\documentclass[a4paper,conference]{IEEEtran}

\usepackage{amsmath}

\DeclareMathOperator*{\argmin}{arg\,min}

\usepackage{amssymb}
\usepackage{subfig}
\usepackage{caption}
\usepackage{pdfpages}
\usepackage{multirow}
\usepackage{graphicx}
\usepackage{epstopdf}

\DeclareMathAlphabet\mathbfcal{OMS}{cmsy}{b}{n}

\usepackage{mathtools}    

\usepackage[noadjust]{cite} 

\begin{document}

\title{	Precoding and Spatial Modulation in the Downlink of MU-MIMO Systems}

\author{Azucena Duarte and Raimundo Sampaio Neto }

\maketitle

\begin{abstract}
This work focuses on the downlink communication of a multiuser MIMO system where the base station antennas and the users' receiving antennas are all active, but at each transmission, only a subset of the receive antennas is selected by the base station to receive the information symbols, and the particular chosen subset (pattern) represents part of the information conveyed to the user. In this paper we present a mathematical model for the system and develop expressions that are fairly general and adequate for its analysis. Based on these expressions we propose a procedure to optimize the choice by the ERB of the sets of antenna patterns to be used in the transmissions to the different users, aiming at the maximization of the detection signal-to-noise ratio. Performance results, with and without the optimization procedure, are presented for different scenarios.
\end{abstract}


\section{Introduction}

Spatial Modulation (SM) and Generalized Spatial Modulation (GSM) \cite{DiRenzo11,DiRenzo14,Humadi14} are recent proposals of communication schemes in MIMO systems. In GSM systems, only a subset of the transmit antennas are activated simultaneously at each transmission timeslot and they are used to send symbols belonging to the symbol constellation of the digital modulation employed, and the particular active antenna combination represents part of the transmitted information. This transmission scheme carries advantages over conventional MIMO systems, once it allows the reduction of the RF chains used by the transmitter, and subsequent increase of the energy efficiency, without significant sacrifice of spectral efficiency.

Recent works \cite{Yang11,Stavridis12,Zhang13} focus on systems called PSM (``Preprocessing aided Spatial Modulation") and GPSM (``Generalized Pre-coding aided Spatial Modulation"). These systems, differently from the previous ones, activates all transmit antennas simultaneously to transmit data only to a subset of the receive antennas, which are selected by the transmitter, and the particular chosen subset represents part of the information conveyed to the receiver. Naturally, the implementation of this schemes requires the use of precoders. The above cited works consider communication between one transmitter and one receiver.

The work herein considers the downlink communication of a multiuser GPSM system, where base station and user antennas are all active, but in each transmission, only subsets of the receive antennas of each user receive information symbols. This paper presents a model for this system and develops fairly general expressions suitable for its analysis, and includes the relation between the total transmit energy and energy $E_k$ available for detection of the signals destined to user~$k$. This relation depends on channel matrices of all users and on the pattern set used by the base station in the transmissions. Based on this relation, a pattern set selection procedure is done by the base station, aiming at the maximization of $E_k$, and subsequent maximization of the detection signal-to-noise ratio and minimization of the error probability. Performance results, with and without optimization procedure, are presented for different scenarios, involving number of system users, number of antennas positioned at the users and number of antennas destined to receive information. 

\section{System and signals}

Consider the downlink of a MU-MIMO system with $N_t$ antennas at the base station and $K$ users, each equipped with $N_r$ receiving antennas, where $N_t\geq K N_r$. In this system, all $N_r$ antennas belonging to the same user are active, but at each transmission only a subset of $N_{\textit{iba}}$ antennas are selected by the transmitter to receive information symbols, and the particular selected subset represents part of the information transmitted by the base station to the user.

Let $N_{\textit{iba}}$ ($N_{\textit{iba}}\le N_r$) the total number of combinations containing $N_{\textit{iba}}$ out of $N_r$ antennas is given by
\begin{equation}
\label{(1)}
C_t =\binom {N_r}{N_{\textit{iba}}},
\end{equation}
and the number of information bits that can be represented by different selections (different patterns) is
\begin{equation}
\label{(2)}
k_{\textit{ssk}}=\lfloor \log_{2}\left(C_{t}\right) \rfloor,
\end{equation}
where $\lfloor  x \rfloor$ denotes the greatest integer less than or equal to $x$. If $M$ is the modulation order, then the total number of bits transmitted by the base station is
\begin{equation}
\label{(3)}
R=K\left(k_{\textit{ssk}}+ N_{\textit{iba}} \log_2 (M)\right),\,\,\, \text{\text{bits/channel use}}
\end{equation}
and $N_c=2^{k_{\textit{ssk}}}$ is the number of valid patterns that can be used by the transmitter.

For instance, let $N_r=4$ and $N_{iba}=2$, resulting in $C_t=6$, $k_{\textit{ssk}}=2$ and $N_c=4$. A possible set of $4$ receive antenna combination patterns can be used by the base station to code $2\, \text{bits}$ of information destined to user $k$, and can be represented by:
\begin{equation}
\label{(4)}
\mathbf{Q}_k=\left[\mathbf{q}_1^k,\,\mathbf{q}_2^k,\mathbf{q}_3^k,\,\mathbf{q}_4^k\right] =\begin{bmatrix}
1 & 1&1&0\\
1&0&0&1\\
0&1&0&0\\
0&0&1&1
\end{bmatrix},
\end{equation}
where $\mathbf{q}_1^k$ indicates that at each transmission slot the 2 information symbols are conveyed to antennas 1 and 2 of user $k$, $\mathbf{q}_2^k$ indicates that antennas 1 and 3 will be the recipients of this information, and so forth. Note that as $N_c \le C_{t}$, $L=\binom {C_t}{N_c}$ possible choices exist for the set $\mathbf{Q}_k$. In the considered example $L=15$ choices exist. As will be shown, the appropriate choice of $\mathbf{Q}_k$ can impact system performance.

\subsection{Signal Model}

Let $\mathbf{s}\in\mathbb{C}^{KN_{r} \times 1}$ the vector that contains the $K$ information vectors conveyed to the users:
\begin{equation}
\label{(5)}
\mathbf{s}=\begin{bmatrix}{\mathbf{s}^{1}}^T,\,{\mathbf{s}^{2}}^T,\,\ldots,\,{\mathbf{s}^{K}}^T\end{bmatrix}^{T},
\end{equation} 
where $\mathbf{s}^k,\, k=1,\,2\ldots,K$, contains the information destined to user $k$. The nonzero entries are determined by the position vectors belonging to $\mathbf{Q}_k$, exemplified in \eqref{(4)}, which are occupied by complex symbols, statistically independent, belonging the signal constellation $\mathcal{C}$ of the modulation employed in the system. Statistically independent $\mathbf{s}^k$ vectors are assumed.

For analysis convenience, the vectors $\mathbf{s}^k$ are represented by
\begin{equation}
\label{(6)}
\mathbf{s}^k=\sqrt{E_{k}}\mathbf{D}(\mathbf{q}^k)\dot{\mathbf{s}}^k,
\end{equation}
where $E_{k}$ is the energy of the information symbols destined to user $k$, $\mathbf{D}(\mathbf{z})$ is the diagonal matrix that contains in its diagonal vector $\mathbf{z}$ and ${\mathbf{q}}^k$ is the random vector, statistically independent of $\dot{\mathbf{s}}^k$, with values drawn from the set $\mathbf{Q}_{k}=\left[\mathbf{q}_1^k,\,\mathbf{q}_2^k,\ldots,\mathbf{q}_{N_{c}}^{k}\right]$, with equal probability. In the representation of \eqref{(6)}, $\dot{\mathbf{s}}^k$ contains symbols belonging to $\mathcal{C}$ in all its $N_r$ entries, all zero mean and unit variance. Thus, $\mathbb{E}[\dot{\mathbf{s}}^{k}]=\mathbf{0}$ and $\mathbb{E}\left[\dot{\mathbf{s}}^{k} \dot{\mathbf{s}}^{k^H}\right]=\mathbf{I}_{N_{r}}$.

Vector $\mathbf{x}\in\mathbb{C}^{N_t \times 1}$ containing the elements transmitted by the base station antennas is given by
\begin{equation}
\label{(7)}
\mathbf{x}=\sum_{m=1}^{K}\mathbf{P}^{m}\mathbf{s}^{m}=\sum_{m=1}^{K}\sqrt{E_m}\mathbf{P}^m\mathbf{D}(\mathbf{q}^m)\dot{\mathbf{s}}^m,
\end{equation}
where $\mathbf{P}^{m},\,m=1,\,2,\,\ldots,\,K$, denotes a $N_t \times N_r$ matrix that precodes the data destined to user $m$.

\subsection{Relation of energies}
	The energy spent by the base station for signal transmission is calculated as
	\begin{equation}
	\label{(8)}
	\begin{aligned}
	E_T=\mathbb{E}[\Vert\mathbf{x}\Vert^{2}]=&\text{Tr}\left\{\mathbb{E}[\mathbf{x}\mathbf{x}^H]\right\}\\
	=&\text{Tr}\left\{\sum_{m=1}^{K}\sum_{l=1}^{K}\mathbf{P}^{m}\mathbb{E}\left[\mathbf{s}^{m}{\mathbf{s}^l}^{H}\right]{\mathbf{P}^{l}}^{H}]\right\}\\
	=&\text{Tr}\left\{\sum_{m=1}^{K}\mathbf{P}^{m}\mathbb{E}\left[\mathbf{s}^{m}{\mathbf{s}^m}^{H}\right]{\mathbf{P}^{m}}^{H}\right\},
	\end{aligned} 
	\end{equation}
	where $\text{Tr}\{\mathbf{A}\}$ denotes the trace of matrix $\mathbf{A}$.
	
	From \eqref{(6)}, we have 
	\begin{equation}
	\label{(9)}
	\begin{aligned}
	\mathbb{E}\left[\mathbf{s}^{m}\mathbf{s}^{m^H}\right]=&E_{m}\mathbb{E}\left[\mathbf{D}(\mathbf{q}^m)\dot{\mathbf{s}}^{m} \dot{\mathbf{s}}^{m^H}{\mathbf{D}}^{H}(\mathbf{q}^m)\right]\\
	=&E_{m}\mathbb{E}\left[\mathbf{D}(\mathbf{q}^m)\right]\\
	=&E_{m}\mathbf{D}\left(\overline{\mathbf{q}}^{m}\right),
	\end{aligned} 
	\end{equation}	
with $\overline{\mathbf{q}}^{m}=\mathbb{E}\left[\mathbf{q}^{m}\right]$ and the equivalence  $\mathbf{D}^{H}(\mathbf{q}^m)=\mathbf{D}(\mathbf{q}^m)=\mathbf{D}^{2}(\mathbf{q}^m)$ was employed.

Combining (\ref{(8)}) and (\ref{(9)}), we get 
\begin{equation}
\label{(10)}
\begin{aligned}
E_T=&\sum_{m=1}^{K}E_{m}\text{Tr}\left\{\mathbf{P}^{m}\mathbf{D}\left( \overline{\mathbf{q}}^{m}\right)  {\mathbf{P}^{m}}^{H} \right\}\\
   =&\sum_{m=1}^{K}E_{m}\text{Tr}\left\{\mathbf{D}\left( \overline{\mathbf{q}}^{m}\right)  {\mathbf{P}^{m}}^{H} \mathbf{P}^{m}  \right\}.
\end{aligned}
\end{equation}
 
 As $\mathbf{D}\left(\overline{\mathbf{q}}^{m}\right)$ is a diagonal matrix, results that 
 \begin{equation}
 \label{(11)}
 \text{Tr}\left\{\mathbf{D}\left( \overline{\mathbf{q}}^{m}\right)  {\mathbf{P}^{m}}^{H} \mathbf{P}^{m}  \right\}={\overline{\mathbf{q}}^{m}}^{T}\mathbf{g}_{m},
 \end{equation}	
with 
\begin{equation}
\label{(11a)}
\mathbf{g}_{m}=\mathbf{d}\left({\mathbf{P}^{m}}^{H} \mathbf{P}^{m}  \right)=\begin{bmatrix}\Vert \mathbf{p}_1^m\Vert^2,\,\Vert \mathbf{p}_2^m\Vert^2,\ldots,\Vert \mathbf{p}_{N_r}^m\Vert^2
\end{bmatrix}^T,	
\end{equation}
where $\mathbf{d}(\mathbf{A})$ denotes the vector whose components are the elements of the diagonal matrix of $\mathbf{A}$ and $\mathbf{p}_{i}^m,\, m=1,\,2,\ldots,K$, represents the $i$th column of matrix $\mathbf{P}^{m}$. Combining \eqref{(10)} e \eqref{(11)} results that 
\begin{equation}
\label{(12)}
\begin{aligned}
E_T=&\sum_{m=1}^{K}E_{m}\mathbf{g}_{m}^{T}{\overline{\mathbf{q}}^{m}}\\
=&E_{s}\sum_{m=1}^{K}\varepsilon_{m}\mathbf{g}_{m}^{T}{\overline{\mathbf{q}}^{m}}
=E_{s} \gamma,
\end{aligned}
\end{equation}
where $E_s=1/K\sum_{m=1}^{K}E_{m}$ is the average energy of the symbols destined to the users and $\varepsilon_m=E_{m}/E_{s}$. The relation between the energy $E_T$ spent in transmission and the symbol energy destined to user $k$ can be expressed as
\begin{equation}
\label{(13)}
E_k=E_{s} \varepsilon_k=E_{T} \frac{\varepsilon_k}{\gamma},
\end{equation}
with 
\begin{equation}
\label{(14)}
\gamma=\sum_{m=1}^{K} \varepsilon_{m}\mathbf{g}_{m}^{T}{\overline{\mathbf{q}}^{m}}.
\end{equation}
From \eqref{(13)} and \eqref{(14)} becomes evident that for a given energy $E_T$ available at the transmitter, the energy $E_k$ available for user $k$ depends on the precoding matrices of all $K$ users, via $\mathbf{g}_{m},\, m=1,\,2,\ldots,K$, given by \eqref{(11a)}, and on the $K$ sets of patterns selected for transmission conveyed to all $K$ user, via 
\begin{equation}
\label{(15)}
{\overline{\mathbf{q}}^{m}}=\frac{1}{N_c}\sum_{i=1}^{N_c} \mathbf{q}_{i}^{m},\, m=1,\,2,\ldots,K.
\end{equation}

\subsection{Optimized choice of the set of patterns}	\label{subsec:pattern_optimize}

Note from \eqref{(13)} and \eqref{(14)} that the minimization of $\gamma$ by means of the $K$ choices of $\mathbf{Q}_m$ results in the maximization of the energy conveyed to each user. As the parcel of the summation in \eqref{(14)} are all positive and each one is a function of the characteristics associated to a single user, the minimization of $\gamma$ can be carried out by the minimization of the parcels independently. In other words, among all $L$ possible choices for the set $\mathbf{Q}$, the one that results in minimal $\mathbf{g}_{m}^{T}{\overline{\mathbf{q}}}$ must be selected for user $m$, with $\overline{\mathbf{q}}=\frac{1}{N_c}\sum_{i=1}^{N_c} \mathbf{q}_{i}$ and $\mathbf{g}_m$ obtained from \eqref{(11a)}. This optimization procedure will be exemplified in Sec.~\ref{sec:num_results}.

\section{Receivers}

Vector $\mathbf{x}$ in (\ref{(7)}) can be rewritten as
\begin{equation}
\label{(16)}
\mathbf{x}=\mathbf{P}\mathbf{s},
\end{equation}
where $\mathbf{P} \in \mathbb{C}^{ N_t  \times K N_r }$ is given by 
\begin{equation}
\label{(17)}
\mathbf{P}=\begin{bmatrix}
\mathbf{P}^{1}\, \mathbf{P}^{2}\,\ldots\mathbf{P}^{K}
\end{bmatrix}
\end{equation} 
and $\mathbf{s}$ is defined as in \eqref{(5)}. Considering the structure of $\mathbf{P}$ given by \eqref{(17)}, results that $\mathbf{P}^{H}\mathbf{P}$ contains in its main diagonal $K$ submatrices ${\mathbf{P}^{m}}^{H}\mathbf{P}^{m},\, m=1,\,2,\ldots,K$. Then, taking \eqref{(11a)} into consideration, results that the vectors $\mathbf{g}_m,$ that appear \eqref{(14)} are given by 
\begin{equation}
\label{(18)}
\begin{bmatrix}
\mathbf{g}_{1}^T\,
\mathbf{g}_{2}^T \,
\ldots 
\mathbf{g}_{K}^T
\end{bmatrix}^T=
\mathbf{d}\left(\mathbf{P}^{H}\mathbf{P}\right).
\end{equation}

Considering \eqref{(16)}, the vector containing the signal received by the $k$-th user can be expressed by
\begin{equation}
\label{(19)}
\begin{aligned}
\mathbf{y}^{k}=&\mathbf{H}_{k}\mathbf{x}+\mathbf{n}_{k}\\
              =&\mathbf{H}_{k}\mathbf{P}\mathbf{s}+\mathbf{n}_{k};
\end{aligned}
\end{equation}
where $\mathbf{n}_{k}$ is the Gaussian noise vector with components circularly symmetric, zero  mean and covariance matrix $\mathbf{K}_{\mathbf{n}_{k}} = \sigma_{n}^{2}\mathbf{I}_{N_r}$. The vector that represents the signals received by all users has the form
\begin{equation}
\label{(20)}
\mathbf{y}=\begin{bmatrix}
\mathbf{y}^{1} \\
\mathbf{y}^{2} \\
\vdots \\
\mathbf{y}^{K}
\end{bmatrix}=
\mathbf{H}\mathbf{x}+\mathbf{n}=\mathbf{H}\mathbf{P}\mathbf{s}+\mathbf{n},
\end{equation}
where $\mathbf{n}={\begin{bmatrix}
\mathbf{n}_{1}^{T},\,\mathbf{n}_{2}^{T},\ldots,\mathbf{n}_{K}^{T}
\end{bmatrix}}^{T}
$ and matrix $\mathbf{H} \in \mathbb{C}^{K N_r \times N_t}$ is composed by $\mathbf{H}={\begin{bmatrix}
\mathbf{H}_{1}^{T},\,\mathbf{H}_{2}^{T},\ldots,\mathbf{H}_{K}^{T}
\end{bmatrix}}^{T}$, where $\mathbf{H}_{k} \in \mathbb{C}^{N_r \times N_t}$ is the matrix that connects the base station antennas to the antennas of user $k$.

\subsection{Systems with ZF precoding}

The ZF precoding matrix is implemented by the left pseudoinverse of $\mathbf{H}$
\begin{equation}
\label{(21)}
\mathbf{P}=\mathbf{H}^{H}(\mathbf{H}\mathbf{H}^{H})^{-1},
\end{equation} 
resulting in
\begin{equation}
\label{(22)}
\mathbf{y}=\mathbf{H}\mathbf{P}\mathbf{s}+\mathbf{n}=\mathbf{s}+\mathbf{n}.
\end{equation}
Then, considering \eqref{(6)}, the vector received by user $k$ becomes 
\begin{equation}
\label{(23)}
\mathbf{y}^{k}=\mathbf{s}^{k}+\mathbf{n}_{k}=\sqrt{E_k}\mathbf{D}\left(\mathbf{q}^{k}\right){\dot{\mathbf{s}}}^{k}+\mathbf{n}_{k}.
\end{equation}

\subsection{Detection}
In this work, maximum likelihood (ML) detection of vector $\mathbf{s}^k$ was considered. A convenient alternative representation of this vector is given by
\begin{equation}
\label{(24)}
\mathbf{D}\left(\mathbf{q}^k\right){\dot{\mathbf{s}}}^{k}= \mathbf{U}^{k}\mathbf{b}^{k},
\end{equation}
where $\mathbf{b}^k \in \mathbb{C}^{N_{\textit{iba}} \times 1}$ is formed by the symbols belonging to constellation $\mathcal{C}$, all zero mean and unit variance. The $N_r \times N_{\textit{iba}}$ matrix $\mathbf{U}^k$ is a submatrix of identity matrix $\mathbf{I}_{N_r}$, obtained from $\mathbf{q}^k$ according to: if the $l$th component of $\mathbf{q}^k$ is zero, then the $l$th column of $\mathbf{I}_{N_{r}}$ is suppressed $\left(l=1,\,2,\ldots,N_{r} \right)$.
Then, the set of patterns $\mathbf{Q}_k$ corresponds to a set of position matrices
\begin{equation}
\label{(25)}
{\mathbfcal{U}}_k=\begin{bmatrix} 
\mathbf{U}_{1}^{k}\, \mathbf{U}_{2}^{k}\, \ldots \mathbf{U}_{N_c}^{k}
\end{bmatrix}
\end{equation}
and ML detector, that decides over the information symbols and corresponding positions in the received vector, can be expressed as 
\begin{equation}
\label{(26)}
\left(\hat{ \mathbf{U}}^{k},\,\hat{ \mathbf{b}}^{k}\right)=
\argmin_{\substack{\setlength{\jot}{-0.8\baselineskip}\begin{aligned}
	\scriptstyle \mathbf{U} &\in \scriptstyle \mathbfcal{U}_{k} \\ 
	 \scriptstyle \mathbf{b} &\in \scriptstyle \mathcal{C}^{N_{\textit{iba}}}
	 \end{aligned}
	 }
 }
{\Vert{\mathbf{y}^{k} - \sqrt{E_{k}} \mathbf{U} \mathbf{b} \Vert}^{2}}, 
\end{equation}
where $E_k=E_T\frac{\varepsilon_k}{\gamma}$, $\gamma$ is given by (\ref{(14)}), and the vectors $\mathbf{g}_m$ are obtained from
\begin{equation}
\label{(27)}
\begin{bmatrix}
\mathbf{g}_1 \\
\mathbf{g}_2 \\
\vdots \\
\mathbf{g}_K
\end{bmatrix}=
\mathbf{d}\left(\mathbf{P}^{H}\mathbf{P}\right)=\mathbf{d}\left( \left( {\mathbf{H}\mathbf{H}^{H}}\right)^{-1} \right).
\end{equation}

\section{Numerical results} \label{sec:num_results}

In this section performance results, obtained by numerical simulation, are presented and expressed in terms of bit error rate (BER) of system users. The elements of all $K$ channel matrices $\mathbf{H}_k,\,k=1,\,2,\ldots,K$ are modeled as statistically independent complex Gaussian random variables, circularly symmetric with zero mean and unit variance entries. Thus, it is admitted that users experience the same path loss. The influence of the channels in the signal detection of the different users is made explicit by \eqref{(13)}, \eqref{(14)} e \eqref{(27)}.

Performance results are expressed in terms of the ratio
\begin{equation}
\label{(28)}
\textit{SNR}=\frac{E_T}{\sigma_n^2},
\end{equation}
where $E_T$ is the total energy spent in transmission, referred to the reception, and $\sigma_n^2$ is the variance of the noise components in the reception. Then, results from \eqref{(13)} that the signal-to-noise ratio per received bit available at the detector in \eqref{(26)} is 
\begin{equation}
\frac{E_k}{\log_2 (M) \sigma_n^2}=\frac{\textit{SNR}}{\log_2 (M)}\frac{\varepsilon_{k}}{\gamma}.
\end{equation}

The modulation employed in transmission is QPSK ($M=4$) and the transmitter destines equal energy to all users ($\varepsilon_{k}=1$).

\begin{figure}
	\begin{center}
		\setlength{\unitlength}{0.0105in}
		\subfloat[\label{f1a}]{%
			\includegraphics[width=.5\linewidth]{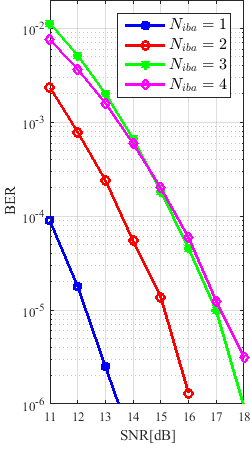}%
		}
		\subfloat[\label{f1b}]{%
			\includegraphics[width=.5\linewidth]{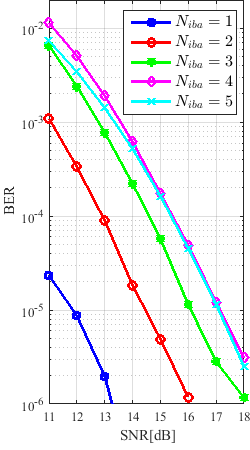}%
		}
	\end{center}
	\caption{\label{f1} BER of GPSM and MIMO systems using ZF precoder. (a) $N_t=8$, $K=1$ and $N_r=4$. (b) $N_t=10$, $K=1$ and $N_r=5$.}
\end{figure}

Figures~\ref{f1a} and \ref{f1b} illustrate, for $N_r=4$ e $N_r=5$, respectively, and $K=1$, system performance for different number of information bearing antennas, $N_{\textit{iba}}$. Note the performance improvement in terms of BER for lower $N_{\textit{iba}}$ values, if $N_{\textit{iba}} < N_r$ is adopted. This performance advantage is achieved at the cost of spectral efficiency reduction, as indicated in Tables~\ref{table1} and \ref{table2}. For the results presented in Figures \ref{f1a} and \ref{f1b}, in the cases that $L> 1$, the set of patterns $\mathbf{Q}$ is fixed and was chosen at random among all possible $L$ choices. The simulation consisted of 1,000 channel matrices realizations with the transmission of 19,200 bits in each realization.

\begin{table}[t]
	\captionsetup{labelsep=newline, justification=centering}
	\centering
	\caption{\label{tabela}\small \textsc{System characteristics $N_t=8$, $N_r=4$, $K=1$.}}
	\label{table1}
	\begin{tabular}{|c|c|c|c|c|}
		\hline
		\textbf{$N_{\textit{iba}}$} & \textbf{$C_t$} & \textbf{$N_c$} & \textbf{$R$} & \textbf{$L$}  \\ \hline\hline
		1                           & 4              & 4              & 4            & 1            \\ \hline
		2                           & 6              & 4              & 6            & 15           \\ \hline
		3                           & 4              & 4              & 8            & 1            \\ \hline
		4                           & 1              & 1              & 8            & 1            \\ \hline
	\end{tabular}
\end{table}

\begin{table}[h]
	\centering
	\captionsetup{labelsep=newline, justification=centering}
	\caption{\small \textsc{System characteristics $N_t=10$, $N_r=5$, $K=1$.}}
	\label{table2}
	\begin{tabular}{|c|c|c|c|c|}
		\hline
		\textbf{$N_{\textit{iba}}$} & \textbf{$C_t$} & \textbf{$N_c$} & \textbf{$R$} & \textbf{$L$} \\ \hline\hline
		1                           & 5              & 4              & 4            & 5            \\ \hline
		2                           & 10              & 8              & 7            & 45           \\ \hline
		3                           & 10              & 8              & 9            & 45            \\ \hline
		4                           & 5              & 4              & 10            & 5            \\ \hline
		5                           & 1              & 1              & 10            & 1            \\ \hline
	\end{tabular}
\end{table}

The following results consider the optimized choice of pattern sets used by the transmitter, by means of the minimization of $\gamma$, according to the procedure proposed in Sec.~\ref{subsec:pattern_optimize}.

\begin{figure}[!t]
	\begin{center}
		\setlength{\unitlength}{0.0105in}
		\includegraphics[width=0.5\textwidth]{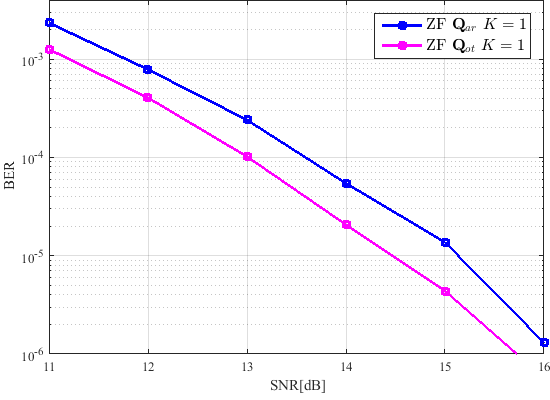}
	\end{center}
	\caption{\label{f2} BER of ZF precoded GPSM system, with randomly chosen $\mathbf{Q}$ ($\mathbf{Q}_{ar}$), and optimized $\mathbf{Q}$ ($\mathbf{Q}_{ot}$). $N_t=8$, $K=1$, $N_r=4$ e $N_{iba}=2$. }
	\label{fig2}
\end{figure}

Figure~\ref{fig2} presents the results for the scenario used in Figure~\ref{f1a} ($K=1,\,N_t=8,\,N_r=4$) with $N_\textit{iba}=2$, , once this is the only value of $N_\textit{iba}$ that permits more than one choice for the set $\mathbf{Q}$. For this example, the $C_t=6$ possible patterns are given by the set $C_t=\{(1,\,2),\,(1,\,3),\,(1,\,4),\,(2,\,3),\,(2,\,4),\,(3,\,4)\}$, where the pair $(i,\,j)$ indicates a pattern with 1's at the $i$th and $j$th positions, respectively, and 0's at the remaining two and the $L=15$ possible choices for the set $\mathbf{Q}$  can be represented by the ordered sets $\mathbf{Q}_1=\{(1,\,2),\,(1,\,3),\,(1,\,4),\,(2,\,3)\},\,
\mathbf{Q}_2=\{(1,\,2),\,(1,\,3),\,(1,\,4),\,(2,\,4)\},\,
\ldots,\,
\mathbf{Q}_{15}=\{(1,\,4),\,(2,\,3),\,(2,\,4),\,(3,\,4)\},
$ with corresponding mean vectors $\overline{\mathbf{q}}$ given by $\overline{\mathbf{q}}_1=1/4\left[3,\,2,\,2,\,1\right]^T,\,\overline{\mathbf{q}}_2=1/4\left[3,\,2,\,1,\,2\right]^T,\,\ldots,\,\overline{\mathbf{q}}_{15}=1/4\left[1,\,2,\,2,\,3\right]^T,$ and the vectors $\mathbf{g}_{m},\,  m=1,\,2\ldots,K$, obtained by \eqref{(27)}.

\begin{figure}[!t]
	\begin{center}
		\setlength{\unitlength}{0.0105in}
		\includegraphics[width=0.5\textwidth]{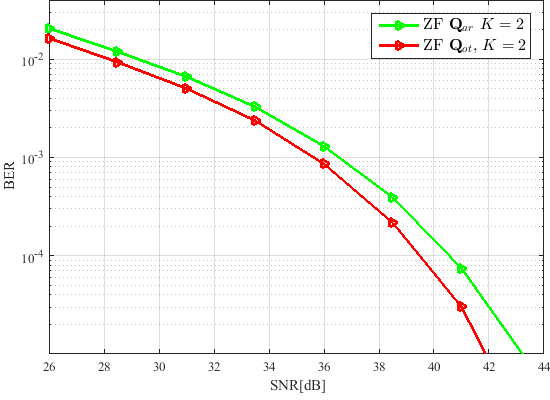}
	\end{center}
	\caption{\label{f3} BER of ZF precoded GPSM system, with randomly chosen $\mathbf{Q}$ ($\mathbf{Q}_{ar}$), and optimized $\mathbf{Q}$ ($\mathbf{Q}_{ot}$). $N_t=8$, $K=2$, $N_r=4$ e $N_{iba}=2$. }
	\label{fig3}
\end{figure}

Results in Figure~\ref{fig3} illustrate a scenario with $K=2$ users ($L=15$). The results of Figures~\ref{fig2} and \ref{fig3} indicate a gain of approximately $1\,\text{dB}$ obtained with the optimization procedure in the considered scenarios.

It is noteworthy that once the choice of the sets $\mathbf{Q}_m$ is done by the transmitter and can vary according to the channel matrix $\mathbf{H}$, the transmitter must inform periodically the users receivers which of the $L$ sets is currently in use, in order to enable the correct signal detection. Results in Figures~\ref{fig2} and \ref{fig3} consider this notification is received free of errors. For comparison purposes, the figures also present the performance obtained with the adoption of a fixed choice for the set $\mathbf{Q}$, known by the users.

A possible way to execute the notification scheme of the set $\mathbf{Q}$ in use, is by means of a frame basis transmission scheme, where at the end of each frame the procedure of choosing the sets $\mathbf{Q}_m,\,m=1,\,2,\ldots,K$ is redone by the transmitter and signals informing the choice made are sent to each user during the notification interval of the following frame. In the case $N_r=4$, $N_{\textit{iba}}=2$ and QPSK modulation, for example, the information of the index of the $L=15$ possible sets can be transmitted using $2$  antennas at the receiver ($4\, \text{bits}$). In order to reduce uncertainty and the possibility of detection error of the notification signal, the antenna pattern used during the periods of notification is known a priori by the receivers. A strategy to further reduce the error probability is to send the same notification information multiple times. The receiver accumulates the received vectors in the notification period and performs detection using the resultant summation vector. In this procedure, if $F$ is the number of repetitions adopted, a signal-to-noise ratio gain of $10 \log_{10} (F)\, \text{dB}$ is obtained.

\begin{figure}[!t]
	\begin{center}
		\setlength{\unitlength}{0.0105in}
		\includegraphics[width=0.5\textwidth]{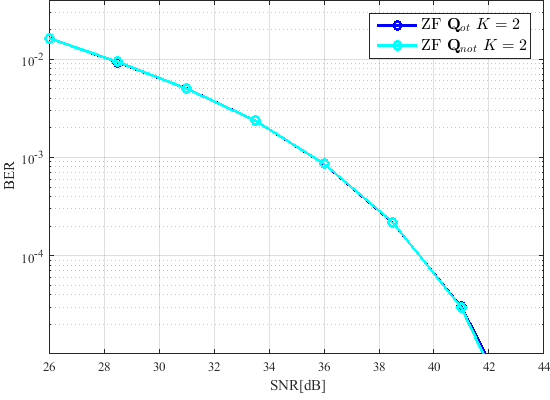}
	\end{center}
	\caption{\label{f4} BER of ZF precoded GPSM system, with optimized $\mathbf{Q}$ known by the receiver ($\mathbf{Q}_{ot}$) and notified to the receiver ($\mathbf{Q}_{not}$). $N_t=8$, $K=2$, $N_r=4$ e $N_{iba}=2$. }
	\label{fig4}
\end{figure}

Figure~\ref{fig4} illustrates the results obtained with the notification strategy described above for the same scenario used in Figure 3. Frames containing $3,200$ information signal vectors ($19,200\,\text{bits}$) to each user were adopted, and $F=10$ repetitions $(40\,\text{bits})$ in the notification interval was adopted. In the simulation, a new sample of the channel matrix was generated at the end of each frame, totaling $1,000$ samples of channel matrices. The accordance between the performance results presented in Figure~\ref{fig4} and those obtained with error-free notification evidences the effectiveness of the proposed notification method.

\section{Conclusion}
This article considered the downlink of GPSM multiuser MIMO systems and developed expressions suitable for the analysis of these systems. Moreover, optimal procedures to determine receive antenna combinations were proposed, along with an effective method of periodic notification of these choices to the users' receivers. Even more pronounced performance gains may be achieved in the optimization procedure from the study of new scenarios and channel models that include, for example, path fading, shadowing and correlation among transmission and reception antennas. These studies are underway.

\section*{Notes}
This paper was published in the Proceedings of the XXXV Brazilian Communications and Signal Processing Symposium~\cite{Azucena17}.

\nocite{*}

\bibliographystyle{IEEEtran}
\bibliography{Reference_ordenadas_profe}

\end{document}